\documentclass[a4paper]{jpconf}

\usepackage[dvips]{color, graphicx}

\newcommand{\Msun}{\textrm{M}_{\odot}}

\begin{document}

\title{Facing the LISA data analysis challenge}

\author{Louis J Rubbo}

\address{Center for Gravitational Wave Physics, Pennsylvania State
  University, University Park, PA 16802 }

\ead{rubbo@gravity.psu.edu}

\date{\today}



\begin{abstract}
  By being the first observatory to survey the source rich low
  frequency region of the gravitational wave spectrum, the Laser
  Interferometer Space Antenna (LISA) will revolutionize our
  understanding of the Cosmos. For the first time we will be able to
  detect the gravitational radiation from millions of galactic
  binaries, the coalescence of two massive black holes, and the
  inspirals of compact objects into massive black holes. The signals
  from multiple sources in each class, and possibly others as well,
  will be simultaneously present in the data. To achieve the enormous
  scientific return possible with LISA, sophisticated data analysis
  techniques must be developed which can mine the complex data in an
  effort to isolate and characterize individual signals. This
  proceedings paper very briefly summarizes the challenges associated
  with analyzing the LISA data, the current state of affairs, and the
  necessary next steps to move forward in addressing the imminent
  challenges.
\end{abstract}


\section{Introduction}

When launched the Laser Interferometer Space Antenna (LISA) will be
the first low frequency ($3 \times 10^{-5} - 0.1$~Hz) gravitational
wave detector~\cite{Danzmann:2003}.  Whenever a new detector is
proposed, especially when it is the first of its kind, there are
plenty of doubts about its capabilities.  LISA is no different.
However, even though the LISA technology is ambitious, its the ability
to analyze LISA data that many think will be the mission's Achilles
heel.

The LISA observatory will return a finite set of time series.  Encoded
within these time series will be the superposition of all
gravitational wave signals received during the mission's observational
run, co-added to a complicated, time dependent instrumental noise
signal.  The goal of LISA data analysis is to coax out an individual
signal from these correlated time series in order to make scientific
inferences about the emitting system or population.  The real
challenge arises because LISA will observe in excess of $10^{8}$
stellar mass galactic binaries, in addition to $0.1 - 10^{5}$ massive
black holes binaries (MBHBs) per year, and up to $10^{3}$ extreme mass
ratio inspirals (EMRIs) per year.  While a daunting task, preliminary
investigations suggest that the LISA data analysis challenge can be
conquered.  This proceedings paper very briefly reviews the
difficulties, achievements, and future directions that the LISA
science community has and will face.

The layout for this paper follows the necessary steps required in
building an analysis routine.  Section~\ref{sec:expectations} briefly
reviews what are our expectations for low frequency sources of
gravitational radiation.  Section~\ref{sec:response} discusses
modeling the detector response and incorporating these models into the
analysis routines.  Section~\ref{sec:analysis} explores what has been
achieved in analyzing simulated data.  The last section points to
future advancements and necessary steps that must be accomplished.

\section{Expectations for astrophysical sources} \label{sec:expectations}

Within the LISA band there are three main classes of sources: a large
galactic population of compact stellar mass binaries; the inspiral,
merger, and ringdown of massive ($10^{4-7}~\Msun$) black hole binaries
-- MBHBs; and the capture, inspiral, and eventual merger of compact
stellar mass objects into massive black holes -- EMRIs.  Each source
class presents unique challenges for data analyst, but as equally
unique is the scientific content that each signal carries.

A number of the greatest challenges associated with LISA data analysis
are concerned with the overwhelming number of stellar mass galactic
binaries.  Initial estimates place the number of individually
resolvable LISA binaries in the several thousands, with millions more
forming an unresolvable background~\cite{Nelemans:2001, Timpano:2006}.
Due to the large orbital periods and low chirp masses associated with
galactic binaries, radiation reaction effects will not drive the
binaries to coalescence during the mission lifetime. In turn, their
signals will be ever present in the detector output.

Conversely, MBHBs are semi-continuous sources.  They begin as a
continuous source during the inspiral phase but eventually fade out
during the ringdown.  Considering the large masses and small orbital
separations, which lead to highly relativistic orbits, the
signal-to-noise ratios (SNRs) for MBHB mergers will make them visible
from throughout the Universe.  The predicted coalescence event rate
ranges from 0.1 per year to a confusion background depending on the
galaxy evolution and black hole growth models~\cite{Berti:2006}.
The wide range in possible event rates hinders the development of data
analysis routines.  An algorithm that searches for the rare MBHB would
be different than one that attempts to isolate an individual signal
within a background.

The predicted detection event rate for EMRIs is $10^{2-3}$ events per
year out to $z \approx 1$ with captures of $10~\Msun$ black holes
accounting for the majority of the rate~\cite{Gair:2004}.  This
estimate was derived using a set of assumptions about analysis
capabilities and a particular astrophysical model.  Nevertheless, it
is evident that a major challenge is detecting EMRIs at large
distances where there is a transition from individual detections to a
possible EMRI background \cite{Barack:2004}.

\section{Forward modeling} \label{sec:response}

As with any measurement in astronomy the telescope acts as a filter
between the incident radiation and the data analyst.  Understanding
the filtering process, sometimes referred to as forward modeling, is
essential in order to extract the full scientific potential hidden in
the data.  Forward modeling plays a significant role for spaceborne
gravitational wave detectors because it is through the continual
orbital motion of the detector that only certain information (e.g. sky
location) becomes encoded in the data.

The LISA mission consists of three identical spacecraft in separate,
slightly eccentric, heliocentric orbits inclined with respect to the
ecliptic plane.  The orbits are chosen such that the constellation
will form an equilateral triangle with a mean spacecraft separation of
$5~\times~10^{6}$~km.  The constellation center will have an orbital
radius of 1~AU and trail the Earth by $20^{\circ}$.  The spacecraft
motion introduces amplitude, frequency, and phase modulations into the
gravitational wave signals.  In modeling LISA's response it is
critical to incorporate these modulations.

An early, complete description for LISA's response was derived by
modifying the response function for terrestrial interferometric
gravitational wave detectors~\cite{Cutler:1998}.  However, since
ground-based detectors operate in the small antenna approximation, the
extension to LISA is only valid for frequencies below $\sim\!10$~mHz,
the point where the gravitational wavelength is on the order of the
detector size.  A higher fidelity response has also been
formulated~\cite{Cornish:2003a}.  Based on this, or a similar
description, multiple open software packages have been developed that
simulate LISA's response to an arbitrary gravitational wave
signal~\cite{Cornish:2003d, Vallisneri:2003, Petiteau:2006}.

\section{Data analysis} \label{sec:analysis}

LISA data analysis is in its early exploratory phase, The typical
strategy undertaken is to develop analysis techniques for each source
class separately with the intent to combine several independent
algorithms to formulate a yet undetermined global analysis procedure.
The following subsections briefly review a few of the highlights in
data analysis developments.

\subsection{Galactic binaries} \label{sub:gb}

Strategies for identifying and characterizing individual bright
galactic binaries include Doppler demodulation
methods~\cite{Hellings:2003, Cornish:2003b}, an iterative subtraction
scheme~\cite{Cornish:2003c}, a tomographic search~\cite{Mohanty:2006},
Markov Chain Monte Carlo (MCMC) approaches~\cite{Umstatter:2005,
Cornish:2005}, and a genetic algorithm~\cite{Crowder:2006}.  While
each technique shows promise, in this limited space we will highlight
the MCMC approach since it appears to be a viable strategy for other
source classes as well.

The central challenge in detecting bright galactic binaries is that in
a small bandwidth ($\Delta f \approx 10^{-6}$~Hz) there may be on the
order of 10 bright binaries with the exact number not known \textit{a
priori}~\cite{Timpano:2006}.  Since each source is described by at
least 7 parameters, the associated parameter space for a small snippet
of the spectrum can be large.  The advantage of a MCMC approach is
that it can quickly (in a comparative sense) explore the parameter
space and return estimates for the parameter values.  In
\cite{Cornish:2005} they demonstrated the ability to detect and
characterize 10 binary signals when the number of systems was a given.
Using a toy model for the signals, \cite{Umstatter:2005} relaxed the
\textit{a priori} assumption concerning the number of systems deriving
its value from the data along with the source parameters.

\subsection{Massive black hole binaries} \label{sub:mbhb}

Only recently has work been done on MBHBs.  The delay may be
attributed to two factors: the large uncertainty in the event rate,
which influences the type of algorithm to design; and the large SNRs
along with their unique ``chirping'' signals implied it would be a
minor exercise to develop a MBHB binary analysis routine.

With this sentiment in mind, a few investigations jumped straight to
the problem of identifying and characterizing a single MBHB signal in
the presence of other signals.  Using a Metropolis-Hastings sampling
with simulated annealing, \cite{Cornish:2006b} was able to isolate a
MBHB signal within a noisy data stream that included a galactic
background.  Separately \cite{Cornish:2006a} and \cite{Wickham:2006}
used a MCMC and a Reversible Jump MCMC (RJMCMC) respectively to
characterize a MBHB signal.  Moreover, \cite{Wickham:2006}
investigated the issue of characterizing a dimmer galactic binary
superimposed by a brighter MBHB and found that it should be possible
to study weaker signals buried beneath brighter signals.

In these isolated examples, the analysis has only focused on the
inspiral phase and, therefore, did not use information from the merger
or ringdown phases.  Also, the MBHB signals where simplified by
ignoring component spins and assuming a circular orbit.  However,
their results are encouraging and suggests future advancements in the
algorithms can account for the neglected effects.

\subsection{Extreme mass ratio inspirals} \label{sub:emri}

Much like with MBHBs, initial analysis algorithms for EMRIs have just
recently been formulated.  However, in contrast to the MBHB case, the
delay for an EMRI analysis algorithm was due to their complicated and
intrinsically weaker signals.  Unlike bright galactic binaries and
MBHBs, the amplitude of a typical EMRI is an order of magnitude below
the instrumental noise.  Only by tracking multiple wave cycles is
enough SNR accumulated to allow a confident detection.

Using scaling arguments, it is possible to show that a standard
template matching routine would require $\sim\!10^{40}$ templates for
an EMRI detection, making it computational prohibitive.  However, this
assumes a fully coherent search.  If instead a hierarchal search is
done by piecing together coherent searches over short observational
periods ($\sim\!3$ weeks) then it may be possible to use a template
based algorithm to search for EMRI signals in the data
\cite{Gair:2004}.

An alternative tactic is to use a time-frequency method in which short
segments of the data time series is Fourier transformed and stacked
together to form a spectrogram~\cite{Wen:2005, Gair:2005}.  By
searching for excess power in the spectrogram it is possible to detect
an EMRI out to distance of $\sim\!2.25$~Gpc, which is about half the
capabilities of the semi-coherent method described previously.

The above analyses are only able to return limited information about
the sources themselves.  In a recent conference proceedings paper
\cite{Stroeer:2006} demonstrated the use of a RJMCMC to characterize a
simplified EMRI signal.  While this analysis only considered leading
order effects to an already approximate EMRI signal, this case study
implies that future improvements may lead to a full EMRI
characterization algorithm.

\section{Concluding remarks} \label{sec:conclusion}

While the LISA data analysis challenge seems difficult, the early
returns indicate that it will be met.  The (RJ)MCMC appears to be a
viable approach for addressing many of the analysis challenges.
However, other techniques may also play a significant role, especially
in the early detection stage where the (RJ)MCMC appears to be limited
in its capabilities.

The next challenges that need to be addressed include processing full
bandwidth (simulated) data without human intervention, cross comparing
existing algorithms, and developing robust routines capable of
analyzing data that contain multiple classes of gravitational wave
sources.  These next steps should be attainable using existing
methods.  However, with LISA data analysis still in its infancy, it is
appropriate and necessary to explore other alternatives in an attempt
to find methods that will maximize our scientific return on the LISA
data.


\ack
This work was supported by the Center for Gravitational Wave Physics
which is supported by the National Science Foundation under
Cooperative Agreement PHY 01-14375.



\section*{References}

\bibliographystyle{unsrt}

\end{document}